# STELLAR ORIGIN OF $^{15}$N-RICH PRESOLAR SIC GRAINS OF TYPE AB:

# SUPERNOVAE WITH EXPLOSIVE HYDROGEN BURNING

Nan Liu[1*], Larry R. Nittler[1], Marco Pignatari[2,3]

Conel M. O'D. Alexander[1], and Jianhua Wang[1]

[1]Department of Terrestrial Magnetism, Carnegie Institution for Science,

Washington, DC 20015, USA

[2]E. A. Milne Centre for Astrophysics, Department of Physics & Mathematics,

University of Hull, HU6 7RX, UK

ABSTRACT

We report C, N, and Si isotopic data for 59 highly $^{13}$C-enriched presolar submicron- to micron-sized SiC grains from the Murchison meteorite, including eight putative nova grains (PNGs) and 29 $^{15}$N-rich ($^{14}$N/$^{15}$N≤Solar) AB grains, and their Mg-Al, S, and Ca-Ti isotope data when available. These 37 grains are enriched in $^{13}$C, $^{15}$N and $^{26}$Al with the PNGs showing more extreme enhancements. The $^{15}$N-rich AB grains show systematically higher $^{26}$Al and $^{30}$Si excesses than the $^{14}$N-rich AB grains. Thus, we propose to divide the AB grains into groups 1 ($^{14}$N/$^{15}$N<Solar) and 2 ($^{14}$N/$^{15}$N≥Solar). For the first time, we have obtained both S and Ti isotopic data for five AB1 grains and one PNG, and found $^{32}$S and/or $^{50}$Ti enhancements. Interestingly, one AB1 grain had the largest $^{32}$S and $^{50}$Ti excesses, strongly suggesting a neutron-capture nucleosynthetic origin of the $^{32}$S excess and thus the initial presence of radiogenic $^{32}$Si ($t_{1/2}$=153 yr). More importantly, we found that the $^{15}$N and $^{26}$Al excesses of AB1 grains form a trend that extends to the region in the N-Al isotope plot occupied by C2 grains, strongly indicating a common stellar origin for both AB1 and C2 grains. Comparison of supernova models with the AB1 and C2 grain data indicates that these grains came from SNe that experienced H ingestion into the He/C zones of their progenitors.

---

[3] NuGrid collaboration, http://www.nugridstars.org.

1. INTRODUCTION

Primitive meteorites contain small amounts of dust with highly anomalous isotopic compositions compared to Solar System materials (e.g., Zinner 2014; Nittler & Ciesla 2016). These presolar grains formed around stars that ended their lives prior to the Solar System formation. So far, tens of thousands of presolar silicon carbide (SiC) grains have been characterized for their C, N, and Si isotopic compositions, based on which five main groups have been defined (Hoppe et al. 1994). The isotopic signatures of mainstream (MS) and X grains definitively link them to low-mass asymptotic giant branch (AGB) stars and supernovae (SNe), respectively. The isotopic compositions of the other types of grains are less diagnostic of their progenitor stars, and of these the AB group grains have the most ambiguous stellar origins (Amari et al. 2001a, hereafter A01a; Nittler & Ciesla 2016).

The AB SiC grains[4] (~5−6% of all SiC) are highly enriched in $^{13}$C ($^{12}$C/$^{13}$C<10) with a wide range of $^{14}$N/$^{15}$N ratios (~50−>10,000). So far, J-type carbon stars and born-again AGB stars have been proposed as sources of AB grains (Alexander 1993; A01a). Hedrosa et al. (2013)

---

[4] A and B grains were originally defined as separate groups with A grains having $^{12}$C/$^{13}$C<3.5 and B grains 3.5≤$^{12}$C/$^{13}$C<10 (Hoppe et al. 1994), with the dividing value of 3.5 being arbitrarily chosen as the equilibrium value of the CNO cycle. Subsequent studies showed no obvious difference between the two groups, which were then combined as a single group, AB (A01a).

reported observational evidence for higher-than-solar $^{14}$N/$^{15}$N ratios in six of seven J-type carbon stars, but the lower-than-solar value for the one exception was a lower limit. State-of-the-art born-again AGB stellar models by Herwig et al. (2011) predict $^{14}$N/$^{15}$N ratios up to ~50,000 in the He intershell region with $^{12}$C/$^{13}$C<10 (Jadhav et al. 2013). Thus, neither scenario seems able to explain the lower-than-solar $^{14}$N/$^{15}$N ratios of ~50% of AB grains, though they cannot be completely ruled out due to poor statistics in the observational measurements and uncertainties in stellar models. Novae are well known stellar explosions in which high-temperature H burning can produce extremely low $^{14}$N/$^{15}$N (0.2−~1000, Jose et al. 2004; Jose & Hernanz 2007) and $^{12}$C/$^{13}$C (<2), and have been proposed as the parent stars of the so-called putative nova grains (PNGs, <0.1% of all SiC) that have extreme $^{13}$C and $^{15}$N enrichments (Amari et al. 2001b). Despite the systematic difference between $^{15}$N-rich AB grains and PNGs, Liu et al. (2016, hereafter L16) pointed out that the isotopic signatures of $^{15}$N-rich AB grains could also reflect nucleosynthetic processes in novae but probably with lower masses relative to PNGs.

Intriguingly, an enhancement in two *p*-process isotopes, $^{92}$Mo and $^{94}$Mo, was found in one AB grain studied by Savina et al. (2003), raising the possibility of a SN origin, the most likely stellar site for *p*-process nucleosynthesis. SN models (Woosley & Heger 2007, hereafter WH07; Pignatari et al. 2013a), however, generally predict extremely high $^{12}$C/$^{13}$C ratios (>10$^5$) in the SN C-rich He/C zone that must have contributed a majority of precursor material to the SiC (e.g., Table 2 of Xu et al. 2015) because thermodynamic equilibrium models require C/O>1 to form SiC. Note that the stable CO molecule could be dissociated in SNe by electrons arising from $^{56}$Co decay, allowing carbonaceous grains to grow even in O>C conditions (e.g., Clayton 2013). However, quasi-equilibrium calculations (Ebel & Grossman 2001) show that although graphite could be stable in O-dominated SN zones, SiC is not stable under such O-rich conditions. Moreover, these zones also have extremely high $^{12}C/^{13}C$ ratios. More importantly, AB and PNG grains show similar Raman characteristics to mainstream SiC grains from low-mass AGB stars after correcting for the isotope-induced peak shifts and broadening (Liu et al. 2017a), suggesting similar condensation environments in their parent stars, i.e., equilibrium condensation with C/O>1. Nittler & Hoppe (2005) reported a presolar SiC grain with the Si and Ti isotopic signature of X grains from SNe, but C and N isotope ratios similar to PNGs, raising the possibility of explosive H-burning occurring in the C-rich He /C zone during a SN explosion. Furthermore, L16 found a subgroup of $^{13}$C- and $^{15}$N-rich presolar SiC, C2 grains (<0.1% of presolar SiC), with neutron-capture isotopic signatures ($^{29,30}$Si, $^{50}$Ti excesses), pointing to a SN origin. Pignatari et al. (2015, hereafter P15) investigated H ingestion into the He/C zone during the pre-SN stage of a massive star's evolution followed by the core-collapse SN explosion. They found that a H-ingestion event before the SN explosion, which is predicted by the stellar model calculations, can switch off the convective mixing process, leaving some H in the He/C zone during the explosion (see P15 and L16 for model details). The combination of the pre-SN H-ingestion event followed by the SN explosive nucleosynthesis can lead to low $^{12}C/^{13}C$ and $^{14}N/^{15}N$, and high $^{26}Al/^{27}Al$ in the H-ingested He/C zone. Thus, SNe should also be considered as possible sources of $^{15}$N-rich AB grains and PNGs.

## 2. METHODS AND RESULTS

The SiC grains in this study were extracted from the Murchison CM2 chondrite using the CsF isolation method described by Nittler & Alexander (2003). Three presolar SiC grain mounts, labeled #1, #2, and #3, were prepared by dispersing grains separated in size by sedimentation (~1 μm) on high purity Au foils from a water suspension. The grains were subsequently pressed into the Au foils with a flat sapphire disk. The $^{13}$C- and $^{15}$N-rich presolar SiC grains analyzed in our previous study were all found on mount #3 (L16). The AB grains and PNGs in this study were nondestructively identified with a Raman microscope by their lower-shifted Raman TO peak (Liu et al. 2017a). Isotopic analyses of C, N, Si, Mg-Al, and Ca-Ti were performed with the NanoSIMS 50L ion microprobe at the Carnegie Institution. Analytical procedures are described in L16. The isotopic data are reported in Table 1 and all the plots with 1σ uncertainties. We relaxed the definition of AB grains ($^{12}C/^{13}C<10$) to $^{12}C/^{13}C<16$ because, for instance, grain M2-A3-G1225 has a $^{12}C/^{13}C$ ratio of 13 but a $^{26}Al/^{27}Al$ ratio of 0.02, an order of magnitude higher than those of MS grains. Among the potential sources of AB grains discussed above, at present only novae and SNe appear likely to produce $^{15}N$-rich N, indicating that it may be useful to re-divide the group into: *group 1 grains with $^{14}N/^{15}N<440$ (solar value),* and *group 2 grains with $^{14}N/^{15}N \geq 440$*. The $^{14}N$-rich and $^{15}N$-rich AB grains are therefore referred to hereafter as AB1 and AB2 grains, respectively.

In addition to AB grains, the rare PNGs and C2 grains, also have $^{12}C/^{13}C <\sim 10$, but have quite restricted $^{14}N/^{15}N$ ratios (~5 to ~50) in contrast to AB grains (~20 to 10,000, Figure 1a). Also, there exist distinctive differences between AB1 and AB2 grains: (1) Presolar SiC with $^{12}C/^{13}C$ lower than the equilibrium CNO cycle value of 3.5 are all $^{15}N$-rich (Table 1); (2) For AB1 grains, the lower the $^{14}N/^{15}N$ ratio, the higher the $^{26}Al/^{27}Al$ ratio, whereas AB2 grains show no correlation in Figure 1a; (3) The linear fits to AB grain Si three-isotope data show that compared to AB2 grains, AB1 grains are more enriched in $^{30}Si$ relative to $^{29}Si$ (Figure 1b). Our N and Al isotopic data for AB1 grains are well correlated and show less scatter than the literature data from A01a, likely indicating less contamination in our study. This is expected because we removed adjacent materials around the grains of interest with a focused ion beam instrument and performed isotope analyses in imaging mode to further exclude contamination during data reduction. Moreover, the size of the O⁻ beam of the NanoSIMS 50L used here is a factor of several smaller than that of the ims-3f instrument used by A01a, which likely further reduced the amount of contamination sampled during our analyses.

Three out of eight AB1 grains and one PNG from this study had negative $\delta^{33,34}$S values outside of 1σ errors (Figure 2a), i.e., $^{32}$S excesses, which could result from the decay of radiogenic $^{32}$Si ($t_{1/2}$=153 yr). Novae are predicted to have normal or positive $\delta^{33}$S compositions (Parikh et al. 2014) and thus cannot explain the $^{32}$S excesses observed in these grains. Pignatari et al. (2013a) first identified $^{32}$S excesses in type C SiC grains as most likely indicating the initial presence of radiogenic $^{32}$Si that was produced by neutron capture in the so-called neutron-burst zone (Meyer et al. 2000) within the He/C zone during a SN explosion. Later, Fujiya et al. (2013) found such radiogenic $^{32}$Si signatures in AB grains and pointed out that the $^{32}$Si excesses could also be produced by the intermediate neutron-capture process (*i*-process, Cowan & Rose 1997) in born-again AGB stars with typical neutron densities of ~$10^{15}$ cm⁻³. Although the AB grains of Fujiya et al. (2013) were not measured for N, we used the $^{26}$Al/$^{27}$Al ratios of these grains to tentatively separate AB1 and AB2 grains based on a dividing value of $^{26}$Al/$^{27}$Al of 0.003 (Figure 1a), shown in Figure 2a. As a matter of fact, two of the three grains from Fujiya et al. (2013) with $^{32}$S excesses are most likely AB1 grains. While the third grain was not measured for its Mg-Al isotopes, it is more enriched in $^{30}$Si relative to $^{29}$Si, increasing the likelihood of it also being an AB1 grain (Figure 1b). In contrast, none of the AB2 grains show any $^{32}$S excesses, and the $^{32}$S excesses therefore seem to be mainly associated with the $^{15}$N-enriched AB1 grains, possibly challenging the born-again AGB scenario suggested by Fujiya et al. (2013). Finally, five out of 12 AB1 grains from our study (Figure 2b) show clear neutron-capture isotopic signatures, i.e., large excesses in $^{50}$Ti ($\delta^{50}$Ti≥~200‰). In contrast, although large $^{32}$S excesses are found in two out of four PNGs, none of the eight PNGs studied so far show any $^{50}$Ti enhancement. All the AB grains and PNGs show quite close-to-solar $\delta^{46,47}$Ti values, indicating close-to-solar initial metallicities for their parent stars (Alexander & Nittler 1999).

3. DISCUSSIONS

Overviews of nova nucleosynthesis in the context of PNG isotopic signatures are given by José et al. (2004) and José & Hernanz (2007). To summarize, although PNGs show isotopic signatures similar to nova nucleosynthetic yields, several fundamental inconsistencies make it problematic to link AB1 grains and PNGs to classical novae: (1) In the context of nova nucleosynthesis, the Si isotope ratios of PNGs indicate that these grains are more likely to originate from ONe novae (1.15−1.35 $M_\odot$), resulting in large excesses in $^{30}$Si relative to $^{29}$Si. Figure 3a, however, shows that the N and Al isotopic data of PNGs can be best explained by CO novae with 1.00−1.15 $M_\odot$ and probably even lower masses, while the $^{15}$N and $^{26}$Al correlation of AB1 grains can be better explained by predictions for a 1.15 $M_\odot$ ONe nova, both of which are completely contrary to the Si isotope constraints. (2) One potential solution to the inconsistency in (1) is that if in the parent CO novae of AB1 grains, the companion low-mass stars had evolved to their AGB phases, the accreted H-rich envelope material onto the white dwarfs would be $^{14}$N- and $^{26}$Al-enriched ($^{14}$N/$^{15}$N>1000, $^{26}$Al/$^{27}$Al=0.001−0.01, Palmerini et al. 2011) compared to the solar-composition envelope adopted in the models. This would lead to a better match with the AB1 grain data by CO nova models, but requires finely tuned timing since the AGB phase only lasts on the order of ~$10^6$ yr. (3) A strong piece of evidence against a nova origin is the neutron-capture isotopic signatures preserved in these grains. In detail, AB1 grains generally show $^{32}$S and $^{50}$Ti excesses, both of which point to neutron irradiation that does not occur in novae (José & Hernanz 2007). Although none of the PNGs analyzed so far show Ti isotopic anomalies indicating neutron capture, two out of four PNGs had large $^{32}$S excesses. The disagreement between isotopic signatures of Si and S measured in PNGs and nova models is illustrated in Figure 3b. Thus, it is unlikely that AB1 grains and at least some of PNGs originated from novae. Finally, stellar models for lower-than 1.00 $M_\odot$ CO novae predict higher-than-solar $^{14}$N/$^{15}$N and could potentially explain AB2 grains that do not show neutron-capture isotopic signatures (Haenencour et al. 2016).

P15 qualitatively investigated the effect of H ingestion in the He/C zone for a 25 $M_\odot$ SN by varying the amount of ingested H from 1.2% (-H), to 0.0024% (-H500) in the He/C zone, and also by adopting two SN shock models: 25T with a peak temperature at the bottom of the He/C zone of 23×10$^8$ K (reproducing the SN shock condition of a 15 $M_\odot$ stellar model), and 25d with a peak temperature of 7×10$^8$ K (the original SN shock condition of a 25 $M_\odot$ model). By comparing with the P15 models, L16 showed that the C, N, and Al isotope ratios of C2 grains can be matched by the 25d models. In addition, the neutron-burst process can still take place in regions of the He/C zone where insufficient amounts of H (≤0.0024%) are ingested. The C2 grain data, therefore, can be explained in the context of SN nucleosynthesis by considering heterogeneous conditions and asymmetries within the He/C zone before the SN shock, possibly triggered by the H-ingestion event.

AB1 grains show well correlated $^{15}$N and $^{26}$Al excesses, pointing to two end-members: $^{15}$N- and $^{26}$Al-rich explosive H-burning products and certain $^{14}$N-rich material with $^{26}$Al/$^{27}$Al<0.001. In spite of differences in the details of the model predictions, existing SN models (WH07; Pignatari et al. 2013a,b, 2015) generally agree that only material from the outermost layers is $^{14}$N-rich but $^{26}$Al-poor within a SN (Lin et al. 2010; Xu et al. 2015). Thus, the model predictions in Figure 4a are shown as mixing lines between explosive H-burning products from the He/C zone with weight-averaged isotopic compositions and material from the outermost zones (8.4 to 13.3 $M_\odot$) in 25T and 25d models. Figure 4a shows that most of the PNGs and two AB1 grains with close-to-terrestrial $^{14}$N/$^{15}$N ratios (if not caused by terrestrial N contamination) are well matched by the mixed ejecta.

On the other hand, both the WH07 and Pignatari models show that the $^{26}$Al/$^{27}$Al ratios of the outermost layers are significantly increased in SN models ≥20 $M_\odot$ relative to those <20 $M_\odot$ (Figures 4b,c), resulting from the loss of the H envelope in the former with the H-burning zone exposed on the surface. Since the mismatch of the models with the C2 and AB1 grain data in Figure 4a is mainly caused by the high $^{26}$Al/$^{27}$Al ratios of materials from the outermost layers, the lower $^{26}$Al/$^{27}$Al ratios of SNe < 20 $M_\odot$ provide a solution. Because the compositions of the explosive H-burning products depend mainly on the peak temperature reached during explosions instead of the stellar mass, we took the same explosive H-burning products in Figure 4a and mixed them with the H-envelope material in the 12 $M_\odot$ WH07 model, shown in Figure 4b. In this case, the N-Al isotope correlation of the AB1 grains is well reproduced by the C-rich 25d and 25d-H5 ejecta, implying that AB1 grains could have come from low-mass core collapse SNe with H ingestion. Alternatively, AB1 grains could also be explained by SNe ≥20 $M_\odot$ if they incorporated a significant amount of materials from the surrounding $^{14}$N-rich H-envelope material with much lowered $^{26}$Al/$^{27}$Al ratio lost during the pre-supernova phase. There is clear observational evidence for interactions between the SN ejecta and the ejected pre-supernova material (e.g., Smith et al. 2015).

More importantly, the N, Al, Si, and Ti isotopic compositions of C2 and AB1 grains strongly indicate that they share a common origin in SNe. Figure 4b shows that C2 and AB1 grains fall quite close to the same mixing line, with C2 grains lying closer to the explosive H-burning products end-member. Moreover, L16 found a $\delta^{50}$Ti excess of 800±200‰ in one C2 grain with a $\delta^{30}$Si value of ~400‰. In comparison, all the AB1 grains have Si isotope ratios within 100‰ of terrestrial values and $\delta^{50}$Ti excesses up to 200‰, which could be explained by diluting the C2 isotopic signature with roughly four times more material from the outermost H envelope that have normal Si and Ti isotopic compositions. In comparison, most of the PNGs can be better explained by more massive SNe compared to AB1 grains (Figure 4a). Moreover, Figure 4b implies that AB1 grains with closer-to-solar N isotope ratios probably can be matched by C-rich ejecta in SNe that achieved peak temperatures lower than $7 \times 10^8$ K (the value adopted in the 25d models) at the bottom of their He/C zones during explosions, implying higher-mass progenitor stars, lower explosive energies, and/or SN asymmetries.

Differences exist between the WH07 and P15 model predictions (Figures 4c,d), which result from differences in the final stellar structures and the adopted nuclear reaction rates. For instance, P15 used the more recent LUNA rates for $^{14}$N$(p,\gamma)^{15}$O that are lowered by up to 40% below $1.5 \times 10^8$ K (Imbriani et al. 2005) compared to the previously recommended values (Angulo et al. 1999), explaining the higher $^{14}$N/$^{15}$N in the P15 models. In addition, both N and Al ratios in Figures 4c,d are also affected by the amount of surface material lost during the presupernova phase.

Interestingly, grain M1-A7-G897 had the largest $^{32}$S and $^{50}$Ti excesses (Figure 2), strongly supporting the concomitant production of both nuclei by neutron capture in the parent SNe. Our results, therefore, support the initial presence of $^{32}$Si in AB1 grains. This also agrees with the fact that AB1 grains are more enriched in $^{30}$Si relative to $^{29}$Si than AB2 grains as a result of stronger neutron-capture nucleosynthesis. As with previous studies (Fujiya et al. 2013; Hoppe et al. 2012; Xu et al. 2015), the derived $^{32}$Si/$^{28}$Si ratios of the three AB1 grains with $^{32}$S excesses from our study are on the order of $10^{-4}$–$10^{-3}$, which are consistent with the 15 $M_\odot$ SN models of Pignatari et al. (2013a), but are generally lower than those of type C grains. Also, in the 12 $M_\odot$ WH07 SN model, enough $^{32}$Si is made in the C-rich He/C zone to explain the grain data, along with their Si and Ti isotope ratios (Figure 5 of L16). In contrast, PNGs show large $^{32}$S excesses but no measurable $^{50}$Ti excesses. If a large amount of Ti from the neutron-burst zone within the He/C zone were removed by condensing TiC, for example, before mixing into the envelope, the decoupled $^{32}$S and $^{50}$Ti correlation of PNGs could be explained. We, however, do not favor this explanation, because Liu et al. (2017b) recently found well correlated Si and Ti isotope ratios in X grains from SNe, suggesting insignificant amounts of fractionation between Si and Ti prior to mixing of ejecta from different zones. The correlated $^{32}$S and $^{50}$Ti excesses in grain M1-A7-G897 also seem not to support this scenario. On the other hand, the decoupled $^{32}$S and $^{50}$Ti excesses of PNGs could be explained if $^{32}$S excesses were incorporated into the grains as $^{32}$S made by alpha capture (Figure 8 of L16).

## 4. CONCLUSIONS

Multi-element isotopic data clearly show distinctive differences between $^{14}$N-rich and $^{15}$N-rich AB grains. We thus propose to divide AB grains into AB1 grains with $^{14}$N/$^{15}$N<440 and AB2 grains with $^{14}$N/$^{15}$N≥440. Detailed comparisons of the new isotope data on these grains with state-of-the-art nucleosynthetic calculations show fundamental problems in linking AB1 grains and PNGs to classical novae. In contrast, there is a satisfactory agreement between the AB1, PNG, and C2 grain data and models for SNe with H ingestion from the envelope into the He/C zone prior to a SN explosion. Interestingly, one AB1 grain with the largest S and Ti isotopic anomalies found in this study, for the first time, shows concomitant overproduction of $^{32}$S and

$^{50}$Ti in the parent star, providing additional evidence to the initial presence of radiogenic $^{32}$Si in presolar grains.

Acknowledgements: This work was supported by NASA's Cosmochemistry program (grant NNX10AI63G to LRN) and benefited from the National Science Foundation under Grant No. PHY-1430152 (JINA Center for the Evolution of the Elements). MP acknowledges significant support to NuGrid from grants PHY 02-16783 and PHY 09-22648 (Joint Institute for Nuclear Astrophysics, JINA), NSF grant PHY-1430152 (JINA Center for the Evolution of the Elements) and EU MIRG-CT-2006-046520, and from SNF (Switzerland). MP also thanks services of the Canadian Advanced Network for Astronomy Research (CANFAR), which in turn is supported by CANARIE, Compute Canada, University of Victoria, the National Research Council of Canada, and the Canadian Space Agency, and ongoing resource allocations on the University of Hulls High Performance Computing Facility viper.

Figure Captions

**Figure 1.** The isotopic compositions of N, Al, and Si of SiC AB grains and PNGs from this study are compared to literature data (Amari et al. 2001a,b; Nittler & Hoppe 2005; L16). For $^{14}$N/$^{15}$N, the protosolar value of 441±5 reported by Marty et al. (2011) is shown as a dash-dotted line for reference. The dashed lines represent terrestrial isotopic compositions. AB grains from this study and A01a were used for the linear fits. $\delta^i$Si is defined as $[(^i\text{Si}/^{28}\text{Si})_{grain}/(^i\text{Si}/^{28}\text{Si})_{std}-1]\times1000$, where $^i$Si denotes $^{29}$Si or $^{30}$Si, and $(^i\text{Si}/^{28}\text{Si})_{grain}$ and $(^i\text{Si}/^{28}\text{Si})_{std}$ represent the corresponding isotope ratios measured in a grain and the SiC standard, respectively.

**Figure 2.** The isotopic compositions of S and Ti of SiC AB grains and PNGs from this study are compared to literature data.

**Figure 3.** Plots of $^{26}$Al/$^{27}$Al versus $^{14}$N/$^{15}$N and $\delta^{34}$S versus $\delta^{30}$Si for presolar SiC grains and nova models. AB grains and PNGs from this study and L16 are compared to the predicted compositions of CO and ONe nova ejecta (José & Hernanz 2007). The model predictions are shown as mixing lines between pure bulk nova ejecta and material of solar composition. For each nova model, the number in parenthesis refers to the percentage of H-rich material accreted onto the white dwarf. The CO nova models predict quite close-to-solar Si and S isotope ratios and therefore cannot be seen in (b).

**Figure 4.** *Upper panel*: The same set of grain data in Figure 3 are compared to the predicted compositions of SN ejecta for N and Al isotope ratios. The model predictions are shown as mixing lines (lines with symbols represent C>O and lines represent C<O) between pure explosive H-burning ejecta and (a) materials from the corresponding outermost layers in the 25d and 25T models, (b) the H-envelope material in the 12 $M_\odot$ WH07 SN model. *Bottom panel*: Plots of model predictions for $^{26}$Al/$^{27}$Al and $^{14}$N/$^{15}$N ratios in the outermost layers versus the initial stellar mass of a SN.

**Table 1.** Isotopic Data of AB Grains and PNGs

## REFERENCES

Alexander, C. M. O'D. & Nittler, L. R. 1999, ApJ, 519, 222
Alexander C. M. O'D. 1993, 78th Annual Meeting of the Meteoritical Society 57: #2869.
Angulo, C., Arnould, M., Rayet, M., et al. 1999, Nuclear Physics A, 656, 3
Amari, S., Nittler, L. R., & Zinner, E. 2001a, ApJ, 559, 463
Amari, S., Gao, X., Nittler, L. R., & Zinner, E. 2001b, ApJ, 551, 1065
Clayton, D. D. 2013, ApJ, 762, 5
Cowan, J.J., & Rose, W.K. 1977, ApJ, 217, 51
Ebel, D. S. & Grossman, L. GCA, 65, 469
Fujiya, W., Hoppe, P., Zinner, E., Pignatari, M., & Herwig, F. 2013, ApJL, 776, L29

Haenecour, P., Floss, C., José, J., et al. 2016, ApJ, 825, #2

Hedrosa, R. P., Abia, C., Busso, M., et al. 2013, ApJL, 768, L11

Herwig, F., Pignatari, M., Woodward, P. R., et al., 2011, ApJ, 727, #89

Hoppe, P., Fujiya, W., & Zinner E., 2012, ApJ, 745, L26

Hoppe, P., Amari, S., Zinner, E., Ireland, T., & Lewis, R. S. 1994, ApJ, 430, 870

Hynes, K. M. & Gyngard, F. 2009, 38 LPI, 1198

Imbriani, G., Costantini, H., Formicola, A., et al. EPJA, 24, 455

Jadhav, M., Pignatari, M., Herwig, F., et al. ApJ, 777: L27

José, J. & Hernanz, M. 2007, M&PS, 42, 1135

José, J., Hernanz, M., Amari, S., Lodders, K., & Zinner, E. 2004, ApJ, 612, 414

Karakas, A. & Lattanzio, J.C. 2014, PASA, 31, #030

Lin, Y., Gyngard, F., & Zinner, E. 2010, ApJ, 709, 1157

Liu, N., Andrew, S., Nittler, R. L., et al. 2017a, M&PS, in revision

Liu, N., Nittler, L. R., Alexander, C. M. O'D., & Wang, J. 2017b, 47 LPI, #2331

Liu, N., Nittler, L. R., Alexander, C. M. O'D., et al. 2016, ApJ, 820, #140

Lodders, K. & Fegley, B. 1995, Meteoritics, 30, 661

Marty, B., Chaussidon, M., Wiens, R. C., Jurewicz, A. J. G., & Burnett, D. S. 2011, Sci, 332, 1533

Meyer, B.S., Clayton, D. D., & The L.-S. 2000, ApJ, 540, L49

Nittler, L. R., & Ciesla, F. 2016, ARAA, 54, 53

Nittler, L. R., & Hoppe, P. 2005, ApJ, 631, L89

Nittler, L. R., & Alexander, C. M. O'D. 2003, GCA, 67, 4961

Palmerini, S., Cognata, M. La, Cristallo, S., & Busso, M. 2011, ApJ, 729, #3

Parikh, A., Wimmer, K., Faestermann, T., et al. 2014, Phys. Lett. B, 737, 314

Pignatari, M. Zinner, E., Hoppe, P., et al. 2015, ApJL, 808, L43

Pignatari, M., Zinner, E., Bertolli, M. G., et al. 2013a, ApJL, 771, L7

Pignatari, M., Wiescher, M., Timmes, F. X., et al. 2013b, ApJL, 767, L22

Smith, N., Mauerhan, J. C., Cenko S. B., et al. 2015 MNRAS, 449, 1876

Woosley, S. E., & Heger, A. 2007, Physics Reports, 442, 269

Xu, Y., Zinner, E., Gallino, R., et al. 2015, ApJ, 799, 156

Zinner, E. 2014, in Treatise on Geochemistry, Vol. 1, ed. A. M. Davis, (Elsevier, Oxford), 181

**Table 1.** Isotopic Data of A+B Grains and PNGs

| Grain | Group | $^{12}C/^{13}C$ | $^{14}N/^{15}N$ | $\delta^{29}Si$ (‰) | $\delta^{30}Si$ (‰) | $^{26}Al/^{27}Al$ (×10$^{-3}$) |
|---|---|---|---|---|---|---|
| M1-A8-G145 | PNG | 4.4±0.01 | 50±2.0 | 31±17 | 157±15 | 69.3±0.9 |
| M2-A1-G410 | PNG | 10.4±0.3 | 38±0.5 | 19±13 | 88±17 | 56.6±1.1 |
| M2-A3-G581 | PNG | 7.8±0.22 | 31±2.4 | 52±11 | 147±12 | 90.2±3.3 |
| M2-A4-G27 | PNG | 2.2±0.05 | 3.8±0.2 | −511±7 | 76±14 | 20.9±7.9 |
| M2-A4-G672 | PNG | 9.6±0.27 | 10±0.8 | −90±18 | 419±28 | 126±4.3 |
| M2-A1-G114 | PNG | 16.2±0.5 | 56±2.0 | 20±25 | 107±35 | |
| M2-A5-G1211 | PNG | 5.9±0.13 | 50±2.5 | −554±11 | −56±35 | |
| M2-A5-G269 | PNG | 8.5±0.19 | 28±1.3 | 7±10 | 59±13 | |
| M1-A4-G386 | AB1 | 6.3±0.01 | 100±2.0 | 8±23 | −29±23 | 6.6±0.14 |
| M1-A4-428 | AB1 | 4.1±0.02 | 195±9.3 | 70±26 | 24±26 | 4.4±0.07 |
| M1-A4-G557 | AB1 | 3.5±0.01 | 144±3.7 | 33±16 | 39±11 | 3.6±0.25 |
| M1-A5-G1424 | AB1 | 3.2±0.01 | 25±0.5 | 51±17 | 50±12 | 16.1±0.19 |
| M1-A7-G894 | AB1 | 11±0.14 | 148±3.8 | −23±26 | 34±31 | 6.8±0.8 |
| M1-A7-G987 | AB1 | 5.0±0.02 | 384±17 | −17±20 | 23±20 | 1.2±0.05 |
| M1-A9-G260 | AB1 | 3.7±0.01 | 64±1.3 | 73±17 | 62±12 | 12.3±0.1 |
| M2-A1-G121 | AB1 | 12±0.34 | 406±27 | −39±14 | 14±18 | 1.3±0.06 |
| M2-A1-G303 | AB1 | 7.1±0.20 | 188±6.3 | −45±11 | −23±14 | 2.8±0.96 |
| M2-A1-G513 | AB1 | 4.5±0.13 | 252±17 | 11±10 | 30±10 | 13.3±0.5 |
| M2-A1-G576 | AB1 | 6.5±0.18 | 155±5.6 | −67±46 | 40±66 | 3.5±1.4 |
| M2-A1-G758 | AB1 | 5.3±0.15 | 410±28 | 131±12 | 87±13 | 1.1±0.11 |
| M2-A2-G567 | AB1 | 4.2±0.09 | 436±65 | 35±15 | 47±18 | 0.9±0.04 |
| M2-A2-G609 | AB1 | 2.8±0.06 | 129±6.8 | −23±12 | −17±15 | 3.3±0.09 |
| M2-A2-G722 | AB1 | 4.1±0.09 | 100±5.0 | 20±14 | 40±18 | 3.0±0.03 |
| M2-A2-G730 | AB1 | 5.8±0.13 | 192±11 | 31±11 | 13±13 | 2.6±0.11 |
| M2-A3-G1225 | AB1 | 13±0.34 | 223±17 | −15±13 | 7±15 | 21.5±2.5 |
| M2-A3-G1637 | AB1 | 9.0±0.14 | 233±19 | −19±12 | 41±13 | 2.3±0.23 |
| M2-A4-G790 | AB1 | 7.6±0.11 | 89±7.0 | −1±13 | 28±14 | 8.3±2.8 |
| M2-A6-G4 | AB1 | 9.8±0.22 | 350±22 | 82±13 | 80±16 | |
| M3-G1134 | AB1 | 5.0±0.13 | 45±1.6 | 67±14 | 77±15 | 8.5±2.7 |
| M3-G1332 | AB1 | 4.0±0.13 | 63±2.1 | 64±10 | 98±11 | 6.3±1.4 |
| M3-G319 | AB1 | 2.7±0.01 | 24±0.4 | 58±15 | 62±17 | 7.6±3.8 |
| M3-G398 | AB1 | 4.3±0.01 | 382±14 | −38±8 | −7±8 | |
| M3-G489 | AB1 | 3.6±0.08 | 26±0.4 | 71±28 | 67±29 | |
| M3-G587 | AB1 | 4.8±0.02 | 300±17 | 171±7 | 141±9 | |
| M3-G1170 | AB1 | 8.0±0.35 | 439±21 | 111±12 | 81±12 | |
| M3-G1607 | AB1 | 4.7±0.02 | 354±9 | 39±10 | 18±13 | |
| M3-G1663 | AB1 | 3.8±0.02 | 394±25 | −43±7 | −11±10 | |
| M1-A9-G353 | AB2 | 9.6±0.03 | 790±83 | 70±16 | 46±10 | |
| M2-A1-G773 | AB2 | 8.4±0.24 | 1862±193 | 109±12 | 45±13 | |
| M2-A1-G807 | AB2 | 10.3±0.3 | 633±30 | 56±11 | 20±12 | |
| M2-A2-G1129 | AB2 | 8.8±0.19 | 1146±94 | 199±9 | 142±10 | |
| M2-A2-G401 | AB2 | 3.8±0.09 | 536±61 | 32±14 | −8±17 | |
| M2-A2-G698 | AB2 | 7.6±0.17 | 651±57 | 74±13 | 45±16 | |

**Table 1.** Continued

| Grain | Group | $^{12}$C/$^{13}$C | $^{14}$N/$^{15}$N | $\delta^{29}$Si (‰) | $\delta^{30}$Si (‰) | $^{26}$Al/$^{27}$Al (×10$^{-3}$) |
|---|---|---|---|---|---|---|
| M2-A2-G729 | AB2 | 12.8±0.3 | 809±111 | 4±14 | 32±18 | |
| M3-GB5 | AB2 | 8.0±0.22 | 4216±432 | 23±9 | 59±10 | |
| M3-G348 | AB2 | 8.6±0.04 | 1401±330 | 34±10 | 62±12 | |
| M3-G406 | AB2 | 10.0±0.04 | 942±70 | −19±9 | 33±10 | |
| M3-G439 | AB2 | 10.0±0.06 | 1503±79 | −34±8 | −13±10 | |
| M3-G447 | AB2 | 11.0±0.06 | 1691±138 | −32±8 | 20±10 | |
| M3-G472 | AB2 | 5.7±0.02 | 620±31 | −24±9 | 0±9 | |
| M3-G756 | AB2 | 11.0±0.08 | 984±119 | 73±12 | 61±15 | |
| M3-G1154 | AB2 | 5.0±0.15 | 523±37 | 11±10 | 58±10 | |
| M3-G1350-4 | AB2 | 11.0±0.28 | 544±20 | 64±9 | 84±9 | |
| M3-G1356 | AB2 | 9.6±0.05 | 1271±236 | 110±9 | 63±11 | |
| M3-G1495 | AB2 | 4.8±0.01 | 917±74 | 7±8 | 26±8 | |
| M3-G1496 | AB2 | 9.3±0.05 | 887±107 | 99±10 | 72±12 | |
| M3-G1507 | AB2 | 8.6±0.04 | 1775±132 | 199±9 | 149±11 | |
| M3-G1669 | AB2 | 8.6±0.04 | 1587±306 | 14±8 | 2±10 | |
| M3-G1693 | AB2 | 10.0±0.06 | 1414±74 | 31±8 | 20±11 | |

| Grain | Group | $\delta^{33}$S (‰) | $\delta^{34}$S (‰) | $\delta^{44}$Ca (‰) | $\delta^{46}$Ti (‰) | $\delta^{47}$Ti (‰) | $\delta^{49}$Ti (‰) | $\delta^{50}$Ti (‰) |
|---|---|---|---|---|---|---|---|---|
| M1-A8-G145 | PNG | −833±167 | −435±131 | 36±39 | 9±23 | −51±19 | 21±20 | 20±39 |
| M2-A1-G410 | PNG | | | 19±82 | 27±15 | −6±13 | 22±15 | −78±15 |
| M2-A3-G581 | PNG | | | | −44±13 | −2±19 | 29±11 | −5±11 |
| M2-A4-G27 | PNG | | | | −48±14 | −81±24 | 47±18 | 49±18 |
| M1-A4-G386 | AB1 | −329±224 | −284±99 | 4±26 | 5±15 | −8±26 | 57±29 | 108±28 |
| M1-A4-428 | AB1 | 86±99 | −59±39 | −26±27 | −42±17 | −61±25 | 10±32 | 206±16 |
| M1-A4-G557 | AB1 | −117±122 | −226±49 | −43±22 | | | | |
| M1-A5-G1424 | AB1 | −69±51 | −84±21 | −79±27 | −14±14 | −48±15 | 6±19 | 207±20 |
| M1-A7-G894 | AB1 | −409±224 | −489±89 | 9±49 | 23±36 | −28±18 | 35±23 | 236±40 |
| M1-A7-G987 | AB1 | −229±447 | −405±167 | −41±40 | −13±19 | −49±10 | 4±14 | 176±49 |
| M1-A9-G260 | AB1 | −374±199 | −276±91 | −70±22 | 3±17 | 21±15 | 59±18 | 61±31 |
| M2-A1-G303 | AB1 | | | −63±54 | | | | |
| M2-A1-G513 | AB1 | | | 157±250 | 62±11 | 22±8 | 41±9 | 67±10 |
| M2-A1-G576 | AB1 | | | −27±13 | | | | |
| M2-A2-G609 | AB1 | | | | 8±28 | −37±58 | −60±12 | −180±7 |
| M2-A2-G730 | AB1 | | | | 23±25 | 25±59 | 60±14 | 32±9 |
| M2-A1-G121 | AB1 | | | −22±138 | | | | |
| M2-A1-G758 | AB1 | | | 88±66 | 18±19 | −37±19 | 79±24 | 182±26 |
| M2-A2-G567 | AB1 | | | | −4±57 | 4±73 | 32±14 | −9±8 |
| M2-A4-G790 | AB1 | | | 43±70 | −12±14 | −11±23 | 4±15 | 45±15 |
| M3-G489 | AB1 | 84±364 | −38±144 | | | | | |

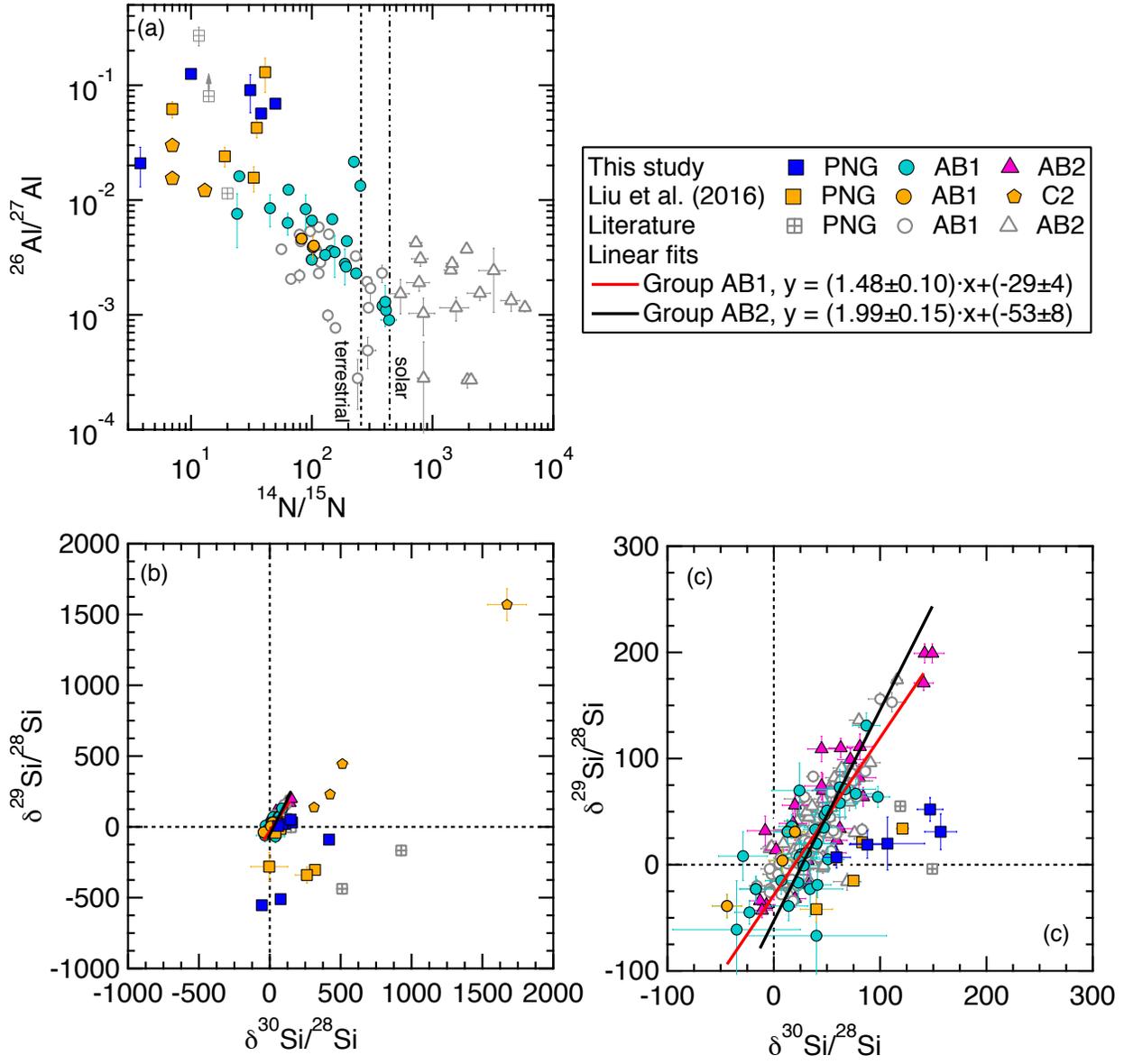

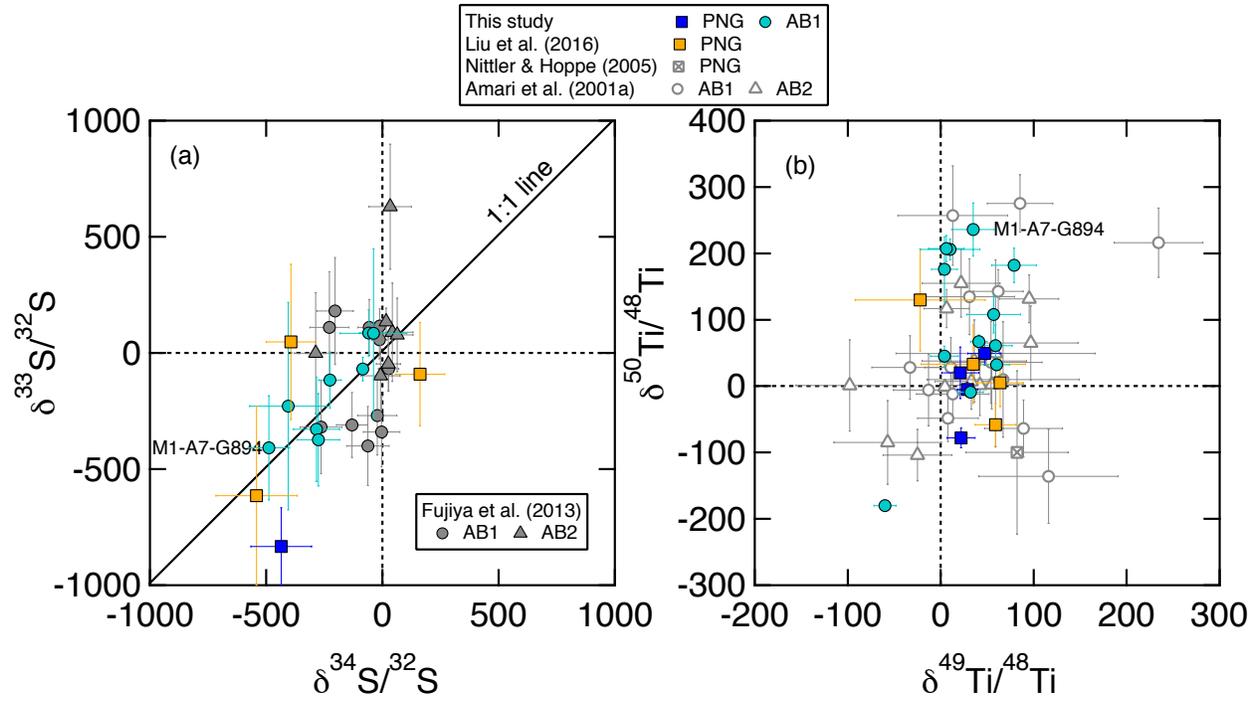

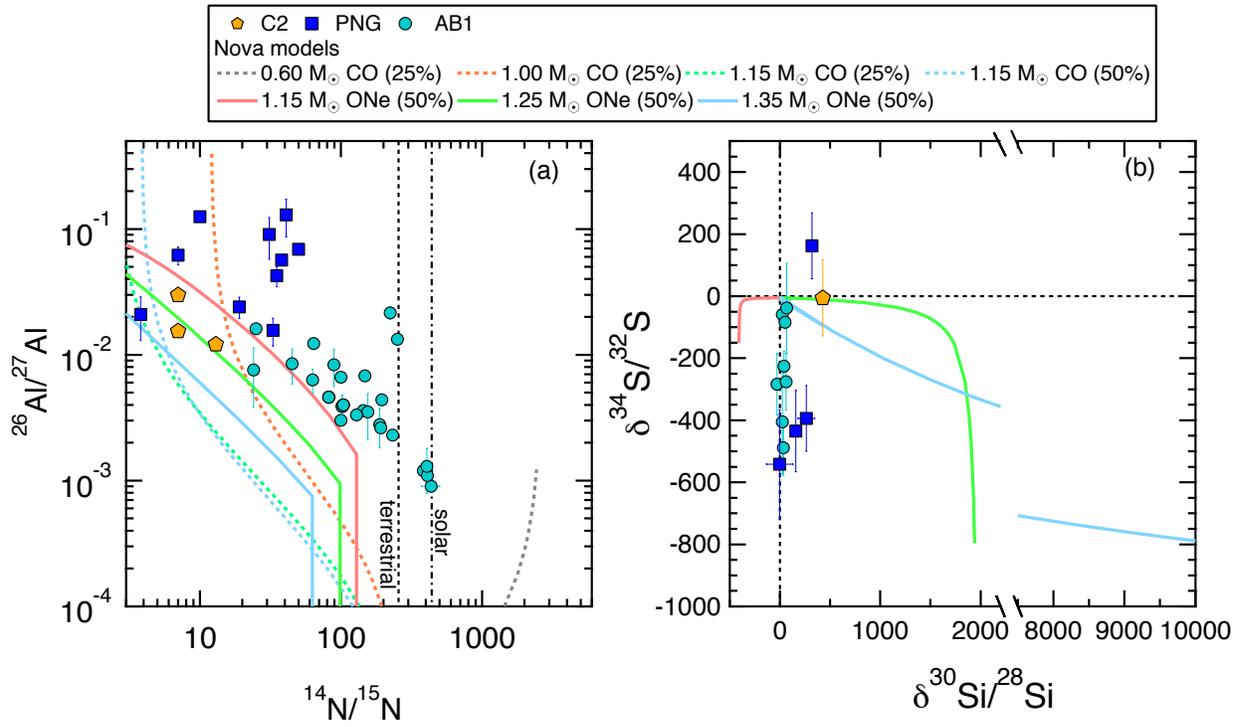

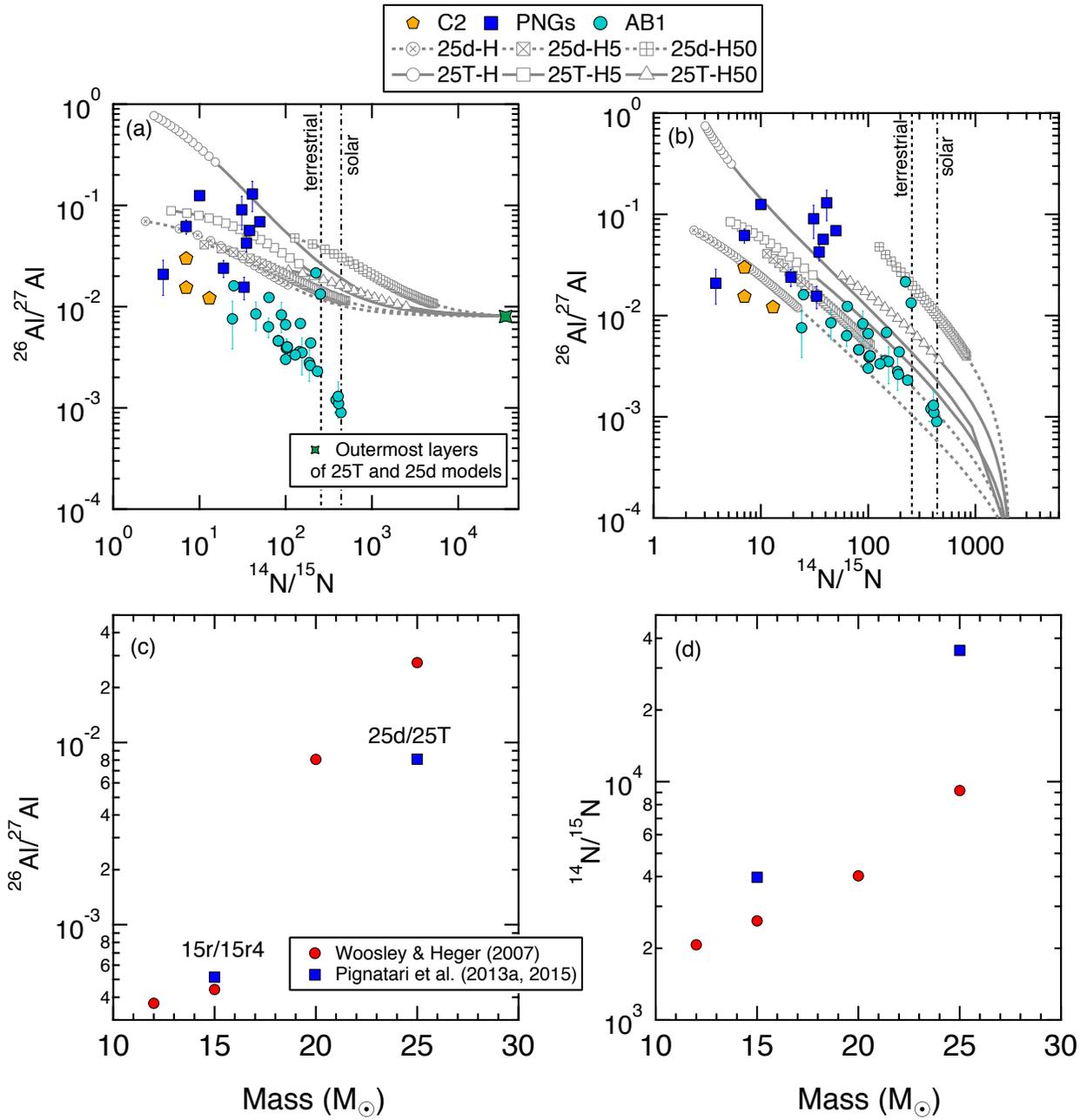